\documentclass[Afour,sageh,times]{sagej}

\usepackage{moreverb,url}

\usepackage[colorlinks,bookmarksopen,bookmarksnumbered,citecolor=red,urlcolor=red]{hyperref}
\usepackage{enumitem}

\newcommand\BibTeX{{\rmfamily B\kern-.05em \textsc{i\kern-.025em b}\kern-.08em
T\kern-.1667em\lower.7ex\hbox{E}\kern-.125emX}}

\def\volumeyear{2016}

\begin{document}

\runninghead{SOME RUNNING HEAD}

\title{The Use of Covariate Adjustment in Randomized Controlled Trials: An Overview}

\author{Kelly Van Lancker\affilnum{1}, Frank Bretz \affilnum{2,3} and Oliver Dukes\affilnum{1}}

\affiliation{\affilnum{1}Department of Applied Mathematics, Computer Science and Statistics, Ghent University, Ghent, Belgium\\
\affilnum{2}Novartis Pharma AG, Basel, Switzerland\\
\affilnum{3}Section for Medical Statistics, Center for Medical Statistics, Informatics, and Intelligent Systems, Medical University of Vienna, Vienna, Austria}

\corrauth{Kelly Van Lancker}

\email{kelly.vanlancker@ugent.be}

\begin{abstract}
There has been a growing interest in covariate adjustment in the analysis of randomized controlled trials in past years. For instance, the U.S. Food and Drug Administration recently issued guidance that emphasizes the importance of distinguishing between conditional and marginal treatment effects. Although these effects coincide in linear models, this is not typically the case in other settings, and this distinction is often overlooked in clinical trial practice.
Considering these developments, this paper provides a review of when and how to utilize covariate adjustment to enhance precision in randomized controlled trials. We describe the differences between conditional and marginal estimands and stress the necessity of aligning statistical analysis methods with the chosen estimand. Additionally, we highlight the potential misalignment of current practices in estimating marginal treatment effects. Instead, we advocate for the utilization of standardization, which can improve efficiency by leveraging the information contained in baseline covariates while remaining robust to model misspecification. Finally, we present practical considerations that have arisen in our respective consultations to further clarify the advantages and limitations of covariate adjustment.
\end{abstract}


\maketitle

\section{Introduction}\label{sec:introduction}
Recent guidance from the U.S. Food and Drug Administration \citep{FDA2021} has led to increased interest in covariate adjustment in the analysis of randomized controlled trials for drugs and biologics. Here, covariate adjustment refers to pre-planned adjustment for prognostic baseline variables (i.e., demographic factors, disease characteristics, or other information collected from patients at the time of randomization) when interest lies in estimating an average treatment effect. This has the potential to improve precision and reduce the required sample size in many clinical trials \citep{tsiatis2008covariate, benkeser2020improving}. While covariate adjustment has been considered already in previous regulatory guidelines \citep{ich1998,EMA2015}, the recent FDA guidance specifically raises the importance of distinguishing between conditional and marginal treatment effects. The distinction between these treatment effects is often overlooked in clinical trial practice because they happen to coincide in linear models. However, covariate adjustment with nonlinear models (e.g., generalized linear models with nonlinear link functions) is often used in the analysis of clinical trial data when the primary outcome of interest is not measured on a continuous scale or is right censored (e.g., binary, ordinal, count, or time-to-event outcomes) and a conditional treatment effect can differ from the marginal treatment effect in these settings \citep{gail1984biased}. Understanding similarities and differences of alternative treatment effects is also in line with the recent ICH E9 addendum on estimands and sensitivity analysis \citep{ich2019}, which calls for a precise description of treatment effect (i.e., estimand) reflecting the clinical question posed by a given clinical trial objective.

In view of the changing regulatory landscape, it therefore becomes important for clinical teams to understand the differences between conditional and marginal estimands and the implication on the statistical analysis approaches that ought to be aligned with the chosen estimand. This raises important questions like: Should a conditional or a marginal estimand be used for a given clinical trial?  Will the selection of covariates lead to different conditional estimands? How to estimate a conditional estimand (e.g., a conditional odds ratio) when fitting a logistic regression with covariates for a binary outcome? How to estimate a marginal estimand (e.g., a marginal odds ratio) and how is it linked to the estimation of a conditional estimand? How is the common practice of fitting a logistic regression (with covariate adjustment) as primary analysis for hypothesis testing related to these estimand considerations?


In light of these questions, we review in this paper when and how to use covariate adjustment to improve precision in randomized controlled trials. Firstly, we describe conditional and marginal treatment effect estimands on the difference, ratio and odds ratio scale. Importantly, these estimands are model-free and their interpretation does not rely on correct specification of models.
Here, the term ``model-free estimand'' refers to an estimand that is not derived from a specific statistical model or a parametric assumption.
Secondly, we touch on the possible misalignment of current practices of treatment effect estimation and discuss in detail the estimation of marginal effect estimands. Thirdly, we address practical considerations that have arisen in our respective consultations on covariate adjustment. 

When discussing the estimation of marginal effect estimands, we provide a standardization estimator, closely related to the one in \cite{FDA2021}, that (1) exploits information in baseline covariates to gain efficiency and (2) is robust to model misspecification. 
Model misspecification refers to the situation where the assumed statistical model used to estimate the treatment of interest does not accurately capture the true underlying data-generating process. It implies that the chosen model does not adequately represent the relationship between the variables under study; e.g., by making incorrect assumptions about the linearity, interaction effects, or distributional properties of the variables.
We also discuss the connection of the standardization estimator with other estimators proposed in the literature.  

\section{Treatment Effect Estimands}\label{sec:estimands}

Let $Y$ denote an outcome of interest (which may be continuous or discrete) and $Z$ a binary, randomised intervention such that $Z=1$ if a patient is assigned to the treatment of interest and $Z=0$ if assigned to the control group (e.g., placebo or an active control). Further, let $X$ denote a collection of variables measured at randomisation. These may include baseline outcome measurements, or (baseline) factors that are known or assumed to be prognostic of the outcome. We assume (for the moment) that there is no loss to follow up, i.e., all randomized patients are assumed to remain in the trial until its end. We also assume that the patients in the trial are a random sample from a larger patient population. 

The ICH E9 addendum emphasizes the need for a precise description of the treatment effect (known as an estimand) that aligns with the clinical question of interest in a clinical trial. According to \cite{ich2019}, an estimand ``\textit{summarises at a population level what the outcomes would be in the same patients under different treatment conditions being compared.}'' In what follows, we focus on the effects of a treatment as assigned, rather than the treatment as taken. This is often referred to as the \textit{intention-to-treat} principle. More formally, we adopt in the following the potential outcomes framework widely used in the causal inference literature to define an estimand of interest \citep{hernan2020, lipkovich2020causal}. Let $Y^z$ denote the potential outcome that would have been observed has someone been assigned to treatment $Z=z$. A given patient has therefore two potential outcomes $Y^1$ and $Y^0$. In practice, one of $Y^1$ or $Y^0$ is unobservable and is called counterfactual. This prohibits learning about treatment effects at an individual patient level, but other causal contrasts can be defined that are of clinical interest and much less ambitious to infer. 

\textit{Marginal} (or population-averaged) causal estimands reflect the effect of treatment in the population defined -- pragmatically speaking -- by the trial's inclusion / exclusion criteria \citep{van2022estimands}. Examples include the average treatment effect on the difference scale
\begin{align}\label{mrd}
E(Y^1-Y^0),
\end{align}
the ratio scale (for positive outcomes)
\begin{align}\label{mrr}
E(Y^1)/E(Y^0),
\end{align}
and the odds ratio scale (for binary outcomes)
\begin{align}\label{mor}
\frac{E(Y^1)/\{1-E(Y^1)\}}{E(Y^0)/\{1-E(Y^0)\}},
\end{align}
where $E$ denotes expectation with respect to the distribution of potential outcomes for the population of interest. In a trial with 
no loss to follow up, each of these effects can be identified by virtue of randomisation. 
Here, identification refers to the translation of a causal estimand into a quantity involving only the observed data so that we can estimate a treatment effect of interest under transparent -- but possibly unverifiable -- assumptions. For example, \eqref{mrd} is identified as $E(Y|Z=1)-E(Y|Z=0)$ under randomization. Intuitively, this is because both treatment arms are balanced due to randomization (at least in large samples) with respect to factors that are prognostic of the outcome. 

One may also consider \textit{conditional} causal estimands where covariates $X$ play an explicit role in the definition of the treatment effect. For example, there may be interest in whether the effect on a certain scale differs between males ($X=0$) and females ($X=1$). Then the causal effect in females on the mean difference scale is \[E(Y^1-Y^0|X=1)\]
and the effect in males can be defined similarly. Alternatively, the effect in females on the odds scale is 
 \[\frac{E(Y^1|X=1)/\{1-E(Y^1|X=1)\}}{E(Y^0|X=1)/\{1-E(Y^0|X=1)\}}.\]

Note that the estimand definitions above are completely model-free.
In practice, treatment effects are often encoded as parameters in, for example, a 
generalised linear model 
\begin{align}\label{eq:glm}
g\{E(Y|Z,X)\}=\beta_0+\beta_1 Z+\beta_2 X,
\end{align}
where $g(\cdot)$ is a pre-specified link function. For example, with continuous outcomes, one may choose $g(\cdot)$ to be the identity link function and fit the model
\begin{align}\label{ancova}
E(Y|Z,X)=\beta_0+\beta_1 Z+\beta_2 X.
\end{align}
This formula encodes the information that there is no interaction between $Z$ and $X$ on the linear scale. This is a statistical modelling assumption not implied by randomisation. Because the estimand is encoded as a parameter in a specific statistical model, we describe the resulting estimand as `model-based'.
If the assumption of no interactions holds, then the model-based estimand $\beta_1$ carries an interpretation as \textit{both} a conditional causal effect $E(Y^1-Y^0|X=x)$ and a marginal causal effect \eqref{mrd}, since they coincide in the case of a linear model without interactions. 

One might also fit a model with an interaction,
\[E(Y|Z,X)=\beta_0+\beta_1 Z+\beta_2 X+\beta_3ZX.\]
Here, $\beta_1$ and $\beta_3$ encode the conditional estimand (on the linear scale), where $\beta_1$ is the effect in those with $X=0$, and $\beta_1+
\beta_3x$ is the effect in those with $X=x$. By including this interaction term in the model, $\beta_1$ and $\beta_3$ typically lose their marginal interpretation, unless $X$ is appropriately centered \citep{ye2022toward}. However, we can use these models to obtain marginal treatment effect estimates by averaging across the observed distribution of baseline covariates, as described in Step 3  in the next section.

For a binary outcome $Y$, the logistic regression model
\begin{align}\label{logit_mod}
logit\{E(Y|Z,X)\}=\beta_0 +\beta_1 Z+\beta_2 X
\end{align}
is commonly used, where $logit(p) = ln(p / (1-p))$ for $p \in (0,1)$.
If this model reflects the truth, then the effect of treatment does not differ between, say, females and males on the relevant scale \citep{vanderweele2014tutorial}. Note that this does not imply that there is no interaction on another scale (e.g., difference or ratio scale). However, unlike in the linear case, $exp(\beta_1)$ would \textit{only} retain an interpretation as a conditional effect,
\begin{align}\label{c_or}
\frac{E(Y^1|X=x)/\{1-E(Y^1|X=x)\}}{E(Y^0|X=x)/\{1-E(Y^0|X=x)\}},
\end{align}
which may differ from the marginal causal odds ratio (\ref{mor}). Therefore, standard practice based on a logistic regression adjusted for covariates typically does not target a marginal causal effect. This phenomenon occurs due to the \textit{non-collapsibility} of the logit link function and is not unique to logistic regression as it also arises in, for example, proportional hazards models \citep{daniel2021making}.  Note that when model \eqref{logit_mod} is misspecified, it may not be clear what standard likelihood-based estimators of $\beta$ are estimating; in particular, they may not generally target either \eqref{mor} or \eqref{c_or}.  

To conclude, a marginal estimand is the treatment effect had all patients in the population been assigned to treatment ($Z=1$) compared with had all patients been assigned to control ($Z=0$). In contrast, a conditional estimand is the treatment effect had a subset of patients with $X=x$ been assigned to treatment compared with had patients with $X=x$ been assigned to control. It is not our intention to wade into the debate as to whether to report marginal or conditional effects for categorical or time-to-event outcomes \citep{Harrell2021Commentary, RemiroAzocar2022}. For an intention-to-treat analysis, marginal estimands can often be inferred by virtue of randomisation alone (assuming no loss to follow up). Further, they arguably generalise more straightforwardly to complex settings with post-randomization events.
However, estimands like (\ref{mrd})-(\ref{mor}) average over the distribution of $X$ in the larger patient population (from which our sample is a random sample), and should be carefully interpreted if the target population where the treatment would be applied differs greatly in terms of $X$ from the patients in the trial \citep{dahabreh2020extending, van2022estimands}. In such cases, model-based conditional effects are sometimes argued to transport better to external populations or generalise to broader populations \citep{vansteelandt2011invited,RemiroAzocar2022}, although this requires that the regression model \eqref{eq:glm} is approximately correct. Estimation of these effects is typically done via fitting regression models using maximum likelihood estimation. Note, however, that the conditional estimand changes if the model changes, e.g. with the inclusion of a prognostic covariate, despite the absence of confounding and large sample size. A marginal estimand remains the same despite the model changing but will typically vary across subgroups of the population. In what follows, we will focus on the estimation of marginal effect estimands as these are usually inferred implicitly in trials if one regresses outcome on treatment alone, or considers the unadjusted difference in means between treatment groups.

\section{Estimation of Marginal Effect Estimands}\label{sec:estimation}
Estimation of marginal effect estimands through an \textit{unadjusted} estimator without covariates is a simple analysis method for the marginal intention-to-treat estimand  and leads to valid inference. This raises the question what can be gained from a more complicated estimation approach that includes covariates. To focus ideas, consider estimating $\mu_1=E(Y^1)$. Its unadjusted estimator, $\sum_{i=1}^nZ_iY_i/\sum_{i=1}^nZ_i$, ignores the information in baseline covariates $X$. The information in the available data is therefore not efficiently used. In contrast, certain \textit{adjusted} estimators familiar from the causal inference literature that use covariate information can lead to a gain in power, without increasing the Type I error rate or introducing bias \citep{tsiatis2008covariate, benkeser2020improving}. Importantly, they rely on randomization and avoid making any modelling assumptions beyond what is assumed for the unadjusted estimator.
In this article, we follow  \citet{FDA2021} and describe one specific approach for covariate adjustment, namely standardization (also referred to as G-computation). It is straightforward to implement and easily explained to collaborators. Alternative covariate-adjusted estimators are discussed in the next section. The proposed standardization estimator for estimating $\mu_1=E(Y^1)$ can be constructed as follows \citep{tsiatis2008covariate}:
\begin{itemize}[left=2em]
    \item[\textbf{Step 1:}] \textbf{Model fitting} \\ Fit a generalized linear regression model (e.g., logistic, linear,...) with a canonical link (e.g., logit, identity,...) via maximum likelihood that regresses the outcome $Y$ on pre-specified baseline covariates $X$ among the treated patients ($Z=1$). This outcome working model should include an intercept term. For example, we could model $E(Y|Z=1, X)$ by $h_1(X;\boldsymbol\gamma)=logit^{-1}(\gamma_0+\gamma_1X)$ for a binary endpoint $Y$. Note that one can also include transformations of the variables in $X$ (e.g., higher order terms and interactions).
    \item[\textbf{Step 2:}] \textbf{Predicting} \\ For each patient, use the fitted outcome working model in Step 1 to compute a prediction of the response under $Z=1$, using the patient's specific baseline covariates. In our example, $h_1(X; \hat{\boldsymbol\gamma})=logit^{-1}(\hat\gamma_0+\hat\gamma_1X)$.
    \item[\textbf{Step 3:}] \textbf{Averaging} \\ Take the average of these predicted responses to obtain an estimator for the average response under $Z=1$. That is, $\hat{\mu}_1=n^{-1}\sum_{i=1}^nh_1(X_i; \hat{\boldsymbol\gamma})$ in our example.
\end{itemize}
Intuitively, although we do not have access to the potential outcomes $Y^1$ for untreated patients, we can make (on average) a guess based on their baseline covariates.
Using a generalized linear model with a canonical link and including an intercept allows us to also use predictions for the treated participants as the mean of their predictions equals the mean of their observed outcomes.

A covariate-adjusted estimator $\hat{\mu}_0$ for $\mu_0$ can be obtained similarly by 
$\hat{\mu}_0=n^{-1}\sum_{i=1}^nh_0(X_i; \hat{\boldsymbol\eta}),$
where the predicted values $ h_0(X_i, \hat{\boldsymbol\eta})$ for each patient are obtained via regression of $Y$ on $X$ among the patients in the control arm ($Z=0$) using a canonical generalized linear working model $h_0(X, \boldsymbol\eta)$ for $E(Y|Z=0, X)$. In the binary endpoint example above, we could then specify $h_0(X, \boldsymbol\eta)=logit^{-1}(\eta_0+\eta_1X)$, in which case $\hat{\boldsymbol\eta}=(\hat\eta_0, \hat\eta_1)'$.
Finally, we construct a covariate-adjusted estimator for \eqref{mrd} as $\hat{\mu}_1-\hat{\mu}_0$, for \eqref{mrr} as $\hat{\mu}_1/\hat{\mu}_0$ and for \eqref{mor} as $\{\hat{\mu}_1/(1-\hat{\mu}_1)\}/\{\hat{\mu}_0/(1-\hat{\mu}_0)\}$. 

To calculate the variance of the estimated treatment effect, it is crucial to take into account that the predictions are derived from specific outcome regression models. Consequently, it is inappropriate to simply compute the sample variance of the difference in predicted (counterfactual) outcomes.
In particular, the variance of the estimators can be obtained via a sandwich estimator using the Delta method \citep{stefanski2002calculus,tsiatis2008covariate,ye2023robust}; for $\hat\mu_1-\hat\mu_0$, it can be calculated as $1/n$ times the sample variances of the values
\begin{align*}
    &\frac{Z_i}{\hat\pi}(Y_i-h_1(X_i; \hat{\boldsymbol\gamma}))+h_1(X_i; \hat{\boldsymbol\gamma}) \\&- \left[\frac{1-Z_i}{1-\hat\pi}(Y_i-h_0(X_i; \hat{\boldsymbol\eta}))+h_0(X_i; \hat{\boldsymbol\eta})\right],
\end{align*}
with $\hat{\pi}=(\sum^n_{i=1}Z_i)/n$ the empirical randomization probability. Alternatively, the variance can be estimated via the non-parametric bootstrap \citep{efron1994introduction}. Although the nonparametric bootstrap is only justified under simple randomization,  \citet{shao2010theory} provide a modification for covariate-adaptive randomization.

Due to randomization, the proposed standardization estimator has the appealing feature that misspecification of the outcome working models $h_1(X, \boldsymbol\gamma)$ and $h_0(X, \boldsymbol\eta)$ does not introduce bias in large samples. We therefore do not need to assume that the link function and the functional form of the covariates (i.e., the algebraic form of the relationship between outcome and covariates) are correctly specified.
In addition, the estimator has the smallest large-sample variance in the class of estimators that are unbiased under randomization alone (including the unadjusted estimator) when the outcome working models are correctly specified. We refer to such estimators as being efficient in large samples.  

The description above is just one specific implementation of the standardization estimator for continuous and binary endpoints. For example, \cite{FDA2021} suggested standardization with one outcome working model that regresses the outcome on treatment assignments and pre-specified baseline covariates. In the description above, we followed \cite{tsiatis2008covariate} and fitted two separate outcome working models as this supports an objective incorporation of baseline covariates by separating the modeling of these relationships from the evaluation of the treatment effect. In addition, such an approach automatically allows for interactions between treatment and covariates, which might lead to additional efficiency benefits. 

Similar standardization estimators have also been introduced in other settings. For example, developments have been made for the  Mann-Whitney estimand \citep{vermeulen2015increasing}, the log-odds ratio for ordinal outcomes \citep{diaz2016enhanced}, restricted mean survival time \citep{chen2001causal,diaz2019improved}, the survival probability difference \citep{wahed2004optimal,lu2008improving}, and the relative risk of time-to-event outcomes; see also \cite{moore2009increasing}, \cite{benkeser2018improved} and \cite{benkeser2019estimating}. 
Recently, \cite{benkeser2020improving} found precision gains for this type of covariate-adjusted estimators in simulated trials for COVID-19 treatments that led to 4-18\% reductions in the required sample size to achieve a desired power. Similarly, a simulation study based on a stroke trial by \cite{VanLancker2022Combining} has shown reductions in sample size of more than 20\% due to covariate adjustment. 

\section{Practical Considerations}
 

Although covariate adjustment offers several benefits as described above, it is not yet routinely used in the analysis of clinical trials. In this section we address practical considerations that have arisen in our respective consultations to further clarify the advantages and limitations of covariate adjustment.

\subsection{How to decide which variables to include?}
This touches on the broader question whether known prognostic covariates (and their functional form) should be pre-specified at the trial design stage or whether they could be selected adaptively based on observed data. Regulatory guidelines usually recommend a pre-specification of covariates and the mathematical form of the covariate-adjusted estimator \citep{FDA1998, EMA2015, FDA2021}. Ideally, the selection is guided by historical trial data in the same disease area and in a similar patient population. 
As a complete pre-specification of the most prognostic variables along with their functional form is a difficult -- if not impossible -- task in many trials, data-adaptive variable and model selection might give one an opportunity to explore covariate-outcome relationships. In particular, the flexibility allowed in the use of modeling methods and expertise makes it possible to make best use of the covariate information to obtain the most efficient estimator. This leads to a tension because for an estimator to be efficient in large samples, the outcome working model must typically be correctly specified, yet choosing this model based on the data is often thought of as complicating or invalidating statistical inference \citep{pocock2002subgroup,tsiatis2008covariate}.

Recent research, however, allows one to use more flexible data-adaptive methods whilst still obtaining valid tests and confidence intervals. This underpins developments  in targeting learning \citep{van2011targeted}, 
and double/de-biased machine learning \citep{chernozhukov2018double}. 
Further, variance of the estimators obtained via sandwich estimators and bootstrap automatically incorporate the uncertainty in the variable selection procedure. 
\citet{williams2021optimizing} provide theoretical and empirical results on incorporating machine learning in the context of randomized trials, along with recommendations for implementation. 

Inference in combination with variable selection / machine learning is a very active area of research. In the future, practice in clinical trials might involve pre-specifying an algorithm to guide the selection of variables in an Statistical Analysis Plan, along with the set of covariates that one starts with \citep{benkeser2021rejoinder, vanlancker2021optimizing}.

\subsection{Should one (only) include variables with imbalance?}
In the medical literature covariates have sometimes been selected for adjustment based on the amount of imbalance across treatment groups, in addition to (or instead of) their association with the outcome \citep{proschan2005adaptive}. This may be motivated by the desire to minimize bias that may arise conditional on the  imbalance \citep{senn1989covariate}. Although intuitive, from the perspective of efficiency it is preferable to prioritize variables that are  prognostic for the outcome. Indeed, such a conditional bias is less of a concern for non-prognostic variables (even if there are imbalances), but adjustment for such covariates can temper precision gains. 
Moreover, even if there happens to be an exact balance in a given sample with respect to a prognostic covariate, unadjusted and adjusted estimators may give identical point estimates but the estimated standard error of the latter tends to be lower than that of the former \citep{proschan2005adaptive}.

\subsection{Can one leverage information from post-baseline covariates?}
One should avoid adjusting for post-baseline covariates in the standardization procedure described in the previous 
section because such variables typically induce selection bias \citep{hernan2004structural}. By conditioning on them, one can induce a spurious association between treatment and unmeasured causes of the outcome; in turn, this can lead to biased estimation of the marginal treatment effect. Furthermore, in the primary analysis of a trial that is not subject to missing data, post-baseline covariates cannot generally be leveraged to improve efficiency. Large sample theory shows that the most efficient estimator of the marginal treatment effect discards information on post-baseline covariate data; see e.g. Lemma 4 in \citet{cheng2021robust}. 

Post-baseline covariates are nevertheless useful in settings where outcomes are subject to missingness or censoring, both in terms of efficiency and for making the missing data assumptions more plausible (see the Question `How to deal with missing data?' below). To avoid selection bias, however, they must be adjusted for via a careful sequential standardization procedure \citep{bang2005doubly}. Post-baseline covariates can also be beneficial in an interim analysis because estimating a treatment effect before the trial end might be viewed as a missing data problem as not all recruited patients will have their primary endpoint observed. \cite{qian2016improving} and \cite{van2020improving} proposed an interim estimator similar to the standardization approach described in the previous section which exploits the information in baseline and post-baseline covariates, without relying on modelling assumptions. Recently, \cite{Tsiatis2022Group} extended this work to treatment effect estimators that account for time-lagged outcomes. These approaches lead to stronger evidence for early stopping than standard, unadjusted approaches, without sacrificing validity or power of the procedure. This is the case even when the adopted outcome working models are misspecified.

\subsection{Can covariate adjustment be harmful?}
The covariate-adjusted standardization estimator is typically efficient in large samples only when the outcome working model is correctly specified. 
However, there exist specific modelling and fitting strategies for the outcome working model (i.e., Step 1 of the algorithm in the previous section) to ensure that the (true) asymptotic variance for the standardization estimator is at least as small as that of the unadjusted estimator, even under model misspecification \citep{rubin2008empirical, diaz2016enhanced, diaz2019improved}. This is most simply implemented in the context of linear models by including all treatment-by-covariate interactions \citep{tsiatis2008covariate,lin2013agnostic,ye2022toward}. 
More generally, Empirical Efficiency Maximisation  is a specific implementation of standardization which is protected against precision loss under misspecification in large samples \citep{rubin2008empirical}. 
Interestingly, an implication of the results in \citet{tsiatis2008covariate} is that covariate adjustment cannot harm large sample efficiency (relative to the unadjusted estimator) in the common scenario with linear outcome models, two treatment groups and 1:1 randomization. Additionally, in this setting there is no efficiency gain from including treatment-by-covariate interactions in the model. 


In finite samples, adjusting for covariates with a very weak or null association with the outcome may not lead to efficiency gains and can even result in a precision loss \citep{kahan2014risks, tackney2022comparison}. It is therefore important that adjustment is performed for covariates that are anticipated to be strongly associated with the outcome. An additional concern is that adjustment may lead to inflated Type I error rates or poor coverage probabilities of confidence intervals when the number of covariates is large relative to the sample size because all variance estimation methods rely on asymptotics.
Although there is no formal rule of thumb, a conservative approach would be to include a small number of key covariates, in line with regulatory guidelines \citep{EMA2015, FDA2021}. Ideally, clinical considerations should be taken into account. Variable selection may also be useful in this context.
Research on how many covariates can safely be adjusted for is ongoing. 

In addition, a small sample size correction can be used for the fact that the variance might be underestimated for finite sample sizes. For example, \cite{tsiatis2008covariate} multiplied the variance estimator by a small-sample correction factor. For the standardization estimator described above,
one can use the correction factor 
$$\frac{(n_0-p_0-1)^{-1}+(n_1-p_1-1)^{-1}}{(n_0-1)^{-1}+(n_1-1)^{-1}},$$
where $n_j (>p_j)$ for $j=0,1$ are the number of patients used to fit the outcome working model in treatment arm $j$ and $p_j$ the numbers of parameters fitted in these models, exclusive of intercepts. 
The (nonparametric) bias-corrected and accelerated \citep[BCa;][]{efron1994introduction} bootstrap has been shown to improve performance for adjusted estimators \citep{benkeser2020improving, VanLancker2022Combining}

\subsection{Are the benefits lost at larger sample sizes?}
Although there is often a concern about the benefits (or rather the loss thereof) in finite sample sizes, sometimes there exists also some confusion about the benefits of covariate adjustment in larger samples. Specifically, the inclusion of covariate adjustments in the trial design is often seen as a desirable but not crucial element, as it is believed that random imbalances tend to diminish with larger sample sizes. As explained in the Question `Should one (only) include variables with imbalance?' above, the latter is not a justification for ignoring baseline covariates. \citet{senn1989covariate} pointed out that covariate imbalance remains a significant concern in large studies, just as it is in small ones. This is because while absolute differences in baseline covariates (i.e., absolute imbalance) may decrease with larger sample sizes, standardized differences, which have an impact on precision, do not decrease.

In fact, the justification for the efficiency gain of covariate adjustment usually derives from large-sample theory. Specifically, the unadjusted and adjusted estimators have different asymptotic distributions. They are both unbiased and asymptotically normal but have different variances, with the variance of the adjusted estimator being generally smaller when the outcome working model is correctly specified. Therefore, one may expect to see improved precision even at large sample sizes.

\subsection{How to decide on the sample size?}
At the design stage of a clinical trial, there is uncertainty about the amount of precision gain and corresponding sample size reduction due to covariate adjustment. 
Determining the required sample size when using covariate-adjusted estimators can be done in at least two ways. One approach is to assume conservatively that covariate adjustment will not lead to a precision gain (in which case any actual precision gains would increase power). Another approach is to consider how much precision can be gained based on external (trial) data when calculating the sample size \citep{li2021estimating}. An incorrect projection of a covariate's prognostic value, however, may still lead to an over- or underpowered future trial. To overcome this, \cite{Tsiatis2022Group} and \cite{VanLancker2022Combining} suggest combining covariate adjustment with information adaptive designs (also known as information monitoring) to take advantage of the precision gain from covariate adjustment. Such an approach would start with setting the sample size at the design stage using one of the two ways described above, and then monitoring the actual precision gain (which is directly related to the statistical information) to determine the timing of the analysis. In particular, this approach will convert the gains into sample size reductions while controlling the Type I error rate and providing the desired power. For instance, when an adjusted estimator is more efficient than an unadjusted estimator, the sample size (and the duration of the trial) will be automatically adapted to the faster information accrual and covariate adjustment will in particular lead to shorter trials.

\subsection{Is there added value from covariate-adaptive randomization?}
Besides adjusting for covariates at the analysis stage, covariates can also be (complementary) accounted for within the randomization process. Examples of covariate-adaptive randomization, which refers to randomization procedures that take baseline covariates into account, include stratified permuted-block randomization, Pocock-Simon's minimization, stratified urn designs, and stratified biased coin designs \citep{sverdlov2015modern}.
Such designs create balance of treatment arms with respect to the stratified variables in order to improve efficiency of the treatment effect estimate.
For example, under the stratified permuted block or biased coin randomization, the standardization estimator described above is asymptotically at least as efficient as the same estimator under simple randomization \citep{wang2021model}.
As a special case, \cite{wang2021model} have shown that the standardization estimator (with separate outcome working models for both arms) has the same asymptotic distribution regardless of whether simple, stratified or biased-coin randomisation is used when using outcome working models including indicators for the randomization strata.
Especially the addition of baseline variables beyond those used for stratified randomization can lead to substantial precision gains \citep{wang2021model}. 
\citet{ye2022toward} have shown that similar results hold for minimisation with a specific implementation of covariate adjustment; 
future research is needed to investigate whether this can be generalised.
Importantly, to obtain the full benefit of covariate-adaptive randomization one should take the randomization procedure into account in the variance estimator \citep{FDA2019, FDA2021}.

\subsection{Is there a connection between different covariate-adjusted estimators?}

A plethora of options is available to adjust for covariates in the estimation of marginal treatment effects besides the standardization approach discussed in the previous section. For example, \citet{tackney2022comparison} compared standardization, Inverse Probability Weighting (IPW), Augmented IPW (AIPW), Targeted Maximum Likelihood Estimation (TMLE) with standard ANalysis of COVAriance (ANCOVA) via simulation. \citet{ye2022toward} recently discussed the ANalysis of HEterogeneous COVAriance (ANHECOVA) procedure, a simple extension of ANCOVA that incorporates treatment-by-covariate interactions; see also \citet{yang2001efficiency} and \citet{lin2013agnostic}. This poses a potentialy challenge to the trialist, who has to choose an approach for the data at hand. 

The competing estimators can be viewed as members  of a general class. As described in \citet{tsiatis2008covariate}, all (reasonable) consistent and asymptotically normal estimators of the marginal risk difference 
can either be exactly written in the form
\begin{align}\label{gen_form}
n^{-1}\sum^n_{i=1} &\frac{Z_iY_i}{\hat{\pi}}-\frac{(1-Z_i)Y_i}{(1-\hat{\pi})}\nonumber\\
&-\frac{Z_i-\hat{\pi}}{\hat{\pi}(1-\hat{\pi})}\left\{(1-\hat{\pi})h_1(X_i;\hat{\boldsymbol\gamma})+\hat{\pi}h_0(X_i;\hat{\boldsymbol\eta})\right\}
\end{align}
 or are approximately equivalent to an expression based on this form. For example, if one chooses $h_1(X;\hat{\boldsymbol\gamma})=h_0(X;\hat{\boldsymbol\eta})=0$, then the above reduces to the unadjusted difference-in-means estimator. 
 
The different proposals for covariate adjustment are thus connected and can in certain cases produce identical point estimates. Suppose we fit a linear model without treatment-by-covariate interactions using ordinary least squares as in \eqref{ancova}; then \eqref{gen_form} reduces exactly to the ANCOVA estimator. If one instead includes a full set of interactions, then \eqref{gen_form} is equal to the ANHECOVA estimator. The IPW estimator described e.g. in \citet{tackney2022comparison} does not explicitly involve modelling the outcome, but instead postulates a logistic regression model for the probability of treatment given covariates. Nevertheless, it follows from \citep{shen2014inverse} that if this model is fit using maximum likelihood, then the IPW and ANHECOVA estimators share the same asymptotic distribution. This is partly because under a simple randomisation scheme, treatment is in truth independent of covariates. If $h_1(X;\hat{\boldsymbol\gamma})$ and $h_0(X;\hat{\boldsymbol\eta})$ are obtained by fitting a generalised linear model with a canonical link function (via maximum likelihood), then \eqref{gen_form} reduces to the standardization estimator described in this paper. AIPW is directly constructed using \eqref{gen_form} and can incorporate both standard parametric and data-adaptive models/estimators for $h_1(X;\hat{\boldsymbol\gamma})$ and $h_0(X;\hat{\boldsymbol\eta})$. It might reduce (exactly or approximately) to one or more of the previously mentioned estimators, based on the choice of outcome model.

When choosing between the different approaches, large-sample theory suggests that the optimal choices of $h_1(X;\hat{\boldsymbol\gamma})$ and $h_0(X;\hat{\boldsymbol\eta})$ are $E(Y|Z=1,X)$ and $E(Y|Z=0,X)$; that is, an efficient estimator will use the true conditional means of the outcome.
Although these are unknown, it motivates estimating them based on a working model that ideally is (approximately) correctly specified. 
In particular, one might tailor a model to a given $Y$ (e.g. a logistic outcome working model for a binary outcome). All estimators mentioned in this section are unbiased in large samples when the outcome working model is misspecified; however, under misspecification the estimator with the smallest large sample variance will depend more delicately on how $\boldsymbol\gamma$ and $\boldsymbol\eta$ are estimated (see also the Question `Can adjustment be harmful?' above). 

\subsection{Is there an added value from `super-covariates'?}
There has been a recent interest in using `super-covariates' as a form of adjustment, where a prognostic score is derived from historical data and used either in addition to or instead of standard baseline covariates. For example, PROgnostic COVariate Adjustment (PROCOVA) \citep{schuler2021increasing} leverages historical data (from control arms of clinical trials and from observational studies) along with prognostic modeling to decrease the uncertainty in treatment effect estimates from randomized controlled trials measuring continuous responses. Being a special case of ANCOVA, PROCOVA preserves the Type I error rate, and is simple to use. In order for it to be fully efficient, the association between outcome and covariates must be equal (or proportional) across the different populations. In addition, it assumes that there are no treatment-by-covariate interactions. Although it will be unbiased when this assumption fails to hold, it may be less efficient than approaches that allow for interactions. In contrast, standardization, AIPW and TMLE are more generally optimal and many of these methods do not require these additional conditions for optimality.
To make best use of the data in the trial, it seems safer to also include (a few) other baseline covariates (e.g., baseline measurement of outcome) instead of using only a single super-covariate.

\subsection{How to deal with missing data?}

In many clinial trials, data may be missing on the baseline covariates or the outcomes. Restricting the analysis to the complete cases, however, can lead to an efficiency loss and possibly biased results. 

Handling missing baseline covariate data in a trial where outcome data is fully observed is often easier than other missing data problems as long as the treatment assignment is independent of the baseline covariates and the missingness mechanism. This is usually expected to hold by virtue of randomisation, allowing the use of simple missing indicator and mean imputation procedures \citep{white2005adjusting, chang2022covariate, zhao2022adjust}. These imputed covariates and missing indicators can then be thought of as a different set of baseline covariates. As covariate adjustment is reliable no matter which pre-specified covariates are used, this approach will remain valid. Nevertheless, less accurate imputation may lead to lower efficiency gain due to the lower prognostic values of the covariates. More complex imputation procedures can also be used, although one should be careful to prevent the imputation model from depending on treatment and the outcome. This is because one must preserve the independence between treatment and covariates in the imputed data. \citet{benkeser2020improving} recommend that baseline covariates with large amounts of missingness should be excluded from the analysis as they will temper the precision gains.

If missing covariates are excluded rather than imputed then randomization still guarantees that the trial can estimate a causal effect, but there are two drawbacks to consider. Firstly, a smaller sample size in the study leads to reduced statistical power. Secondly, the treatment effect can only be generalized to a modified population that includes individuals with non-missing baseline covariates.

When outcome data are missing or time-to-event endpoints are censored, baseline covariates can be utilised not just for efficiency gains, but also to relax missing data assumptions. Specifically, one can assume that outcomes are Missing At Random (MAR) rather than Missing Completely At Random (MCAR), meaning that missingness should be independent of the outcomes conditional on treatment and covariates. To perform an analysis, one can then tweak the aforementioned standardization procedure to make predictions for all patients in the dataset, rather than just for the complete cases. This is sometimes known as regression imputation or single imputation \citep{little2019statistical}. The validity of the resulting treatment effect estimator now relies on a correctly specified outcome working model for the complete cases. Multiple imputation is an alternative approach which relies on the same modelling assumptions \citep{rubin2004multiple, schafer1997analysis}, and uses `Rubin's Rules' to account for uncertainty in the imputations. By also incorporating (in the regression imputation approach) a model for the missing data mechanism, one can construct an estimator that is doubly robust; that is, unbiased if either the outcome model or the missingness model is correctly specified. This can be accomplished by using a weighted outcome regression in Step 1 of the algorithm in the previous section for the complete cases, with weights one divided by the probability of being a complete case (i.e., one minus probability of having a missing outcome) given baseline covariates. These methods can also accommodate time-varying prognostic covariates of the outcome that are associated with missingness under a sequential MAR assumption, although this requires a more involved implementation \citep{bang2005doubly}. 

In the case that both outcome and covariate data are missing, the mean imputation and missing indicator strategies for covariates may no longer suffice as MAR may not hold conditional on these imputed covariates. How to best proceed in the general setting remains a topic for future research.

\subsection{What are the implications for hypothesis testing?}

For the sake of simplicity and to be consistent with current practice, we focus on fitting one model including an indicator for treatment and baseline covariates. Let $\hat\beta=(\hat\beta_0, \hat\beta_1, \hat\beta_2)$ denote the estimated coefficients when fitting a generalised linear model with pre-specified canonical link function \eqref{eq:glm}
using maximum likelihood estimation.
The corresponding standardized estimator of the marginal risk difference, $\hat\mu_1-\hat\mu_0$, is then defined as:
\[n^{-1}\sum_{i=1}^ng^{-1}(\hat\beta_0+\hat\beta_1 +\hat\beta_2 X_i)-n^{-1}\sum_{i=1}^ng^{-1}(\hat\beta_0 +\hat\beta_2 X_i).\]

To test the null hypothesis of no marginal treatment effect $H_0: E(Y^1)=E(Y^0)$, one can either test $\beta_1=0$ or $\mu_1-\mu_0=0$. It turns out that the resulting Wald tests based on the respective estimates $\hat\beta_1$ and $\hat\mu_1-\hat\mu_0$ both control the Type I error rate and are equally powerful in large samples \citep{rosenblum2016matching}. Note that this also holds under arbitrary model misspecification of the outcome (working) model. \cite{rosenblum2016matching} also conducted a simulation study that shows that these results approximately hold for finite samples. 
Thus, the misalignment of current practices for marginal treatment effect estimation does not hold for testing in large samples as long as one does not include interactions. Nevertheless, enhanced standardization estimators (e.g., by fitting separate outcome working models or by including a model for randomization) have the potential for greater efficiency gains when the outcome working models are misspecified.

\subsection{Which software should I use?}
To promote the use of covariate adjustment in practice, it is necessary that easy-to-use, well tested, open-source software is accessible to researchers. 
The R package \textit{RobinCar} includes covariate adjustment for the common outcome types (continuous, discrete, and time-to-event). It includes the standardization approach as well as the ANHECOVA methodology of \citet{ye2022toward}, and can also be used in conjunction with covariate-adaptive randomization schemes.
In terms of a brief survey of some of the available software in R, the package \textit{speff2trial} performs estimation and testing of the treatment effect in a 2-group randomized clinical trial with a quantitative, dichotomous, or right-censored time-to-event endpoint; \textit{drord} implements the efficient covariate-adjusted estimators described in \citet{benkeser2020improving} for establishing the effects of treatments on ordinal outcomes; 
 \textit{survtmle} performs targeted estimation of marginal cumulative incidence for time-to-event endpoints with and without competing risks as described in \citet{benkeser2018improved}; \textit{adjrct}  implements efficient estimators for the restricted mean survival time and survival probability for time-to-event outcomes and the average log odds ratio and Mann-Whitney estimand for ordinal outcomes without proportional hazards and odds assumptions \citep[see][]{vermeulen2015increasing,diaz2016enhanced, diaz2019improved}. 
 \cite{BetzVanLanckerRosenblum} developed tutorials that are meant to help trialists make use of covariate adjustment across the lifespan of a randomized trial, from pre-trial preparations to interim analyses and final reporting of results. The purpose of these tutorials is to emulate practical settings and datasets commonly encountered in real-world scenarios.
 One area for further software development is empirical efficiency maximisation 
\citep{rubin2008empirical}, since these methods may help to assuage concerns regarding efficiency loss in the presence of model misspecification.

\section{Discussion}
In this paper, we discuss the use of covariate adjustment in randomized controlled trials to improve precision, which has become increasingly important due to changing regulatory requirements \citep{ich2019, FDA2021}. We highlight the distinction between conditional and marginal estimands and the need to align statistical analysis approaches with the chosen estimand. We also point out the potential misalignment of current practices for marginal treatment effect estimation. Instead, we advocate using standardization, which can increase efficiency by exploiting the information in baseline covariates while remaining robust to model misspecification. Finally, we offer practical considerations related to covariate adjustment for clinical trialists to keep in mind. Although we have focused on marginal estimands in this paper, we note that similar developments could in principal be made for conditional estimands, along the lines of \citet{vansteelandt2022assumption}; this warrants further study. 

In a recent call toward better practice of covariate adjustment in analyzing randomized clinical trials, \cite{ye2022toward} presented three practical requirements for covariate adjustment via model-assisted approaches: (a) no (asymptotic) efficiency loss compared to an estimator that does not adjust for covariates; (b) applicable to commonly used randomization schemes; and (c) valid inference based on robust standard errors, even under model misspecification. We agree with their considerations and believe that besides ANHECOVA \citep{ye2022toward}, which is most naturally suited to continuous outcomes, the standardization approach discussed in this paper can also satisfy these requirements (see the Questions `Can covariate adjustment be harmful?',  `Is there added value from covariate-adaptive randomization?' and `Is there a connection between different coariate-adjusted estimators?' in the previous section). 

In this article, we mainly focused on the use of covariate adjustment to define causal conditional estimands (e.g., subgroup effects) and to improve the precision when estimating marginal causal effects, as these were the most significant contribution in \citep{FDA2021}. Baseline covariates are also valuable to be used as stratification factors to ensure balance of treatments across covariates.
There  is also a growing recognition among researchers of the importance of accounting for intercurrent events to ensure the validity of results \citep{ich2019}.
In this article we have focused on the intention-to-treat estimand which considers the occurrence of intercurrent events irrelevant in defining the effect of the (assigned) treatment. 
However, additional strategies have been considered which explicitly account for intercurrent events  \citep{ich2019}.
Unlike treatment assignment at baseline, the occurrence of an intercurrent event is not randomized. Many of the estimands referred to in the ICH E9 addendum are therefore only identified under stronger assumptions, even in the absence of loss to follow up. In particular, identification of a causal effect when the intercurrent event forms part of the intervention often requires measurement of baseline as well as post-baseline covariates. 
For example, estimands defined under a hypothetical strategy would usually require measurement of time-varying confounders \citep{olarte2022hypothetical,stensrud2022translating}.
This is similar to observational studies in which  baseline covariates also play a vital role to control for
confounding. By including baseline covariates in the analysis, we can control for these confounding factors and obtain a more accurate estimate of the exposure-outcome relationship.

We believe that incorporating covariate adjustment can enhance the quality and efficiency of clinical trials. To encourage wider adoption of this approach, it would be beneficial to have multiple estimators available within a single R package. Combining this with guidelines on how to specify the considered estimand and covariate-adjusted estimator in a study protocol may prompt statisticians to more carefully consider appropriate estimands and corresponding estimators, ultimately leading to an increased and more thoughtful use of covariate adjustment in clinical trials.

\begin{acks}
We would like to thank Mouna Akacha, Mark Baillie, Bj\"orn Bornkamp, Christopher Jennison, Hege Michiels, Dan Rubin, Stijn Vansteelandt, Jiawei Wei, and Ting Ye for helpful comments on an earlier version of this paper.
\end{acks}

\bibliographystyle{SageH}
\bibliography{Bibliography-MM-MC}

\end{document}